\documentclass[fleqn,twoside]{article}
\usepackage{espcrc2}

\usepackage{graphicx}
\usepackage[figuresright]{rotating}

\newcommand{\AmS}{{\protect\the\textfont2
  A\kern-.1667em\lower.5ex\hbox{M}\kern-.125emS}}

\title{NEUTRINO MASS-AN OVERVIEW}

\author{R. N. Mohapatra\address[MCSD]{Department of Physics, University of
Maryland, College Park, MD-20742, USA}%
        \thanks{Work supported by the National Science Foundation Grant
No. PHY-0099544}}
       
\begin{document}

\begin{abstract}
A brief overview of the present status of neutrino mass physics is given.
\vspace{1pc}
\end{abstract}

\maketitle

\section{INTRODUCTION}
       There is now convincing evidence from solar, atmospheric,
accelerator as well as reactor neutrino studies that neutrinos, long
thought to be massless, are indeed massive and like the quarks, they mix
among
themselves leading to the phenomenon of neutrino oscillations. Since the
standard
model predicts that neutrinos are massless, this is the first conclusive 
evidence for physics beyond the standard model and as such, has led to
a new and important phase in the exploration of physics beyond the
TeV scale. In this talk, I will present a brief overview of
our present understanding of neutrino properties, what the recent
discoveries have taught about new physics and where
we go from here.

Throughtout this report, we will use the notation, where the flavor or
weak eigenstates are denoted by $\nu_{\alpha}$ (with $\alpha~=~e, {\mu},
{\tau},\cdot\cdot\cdot$) and they 
are expressed in terms of the
mass eigenstates $\nu_{i}$ ($i=1, 2, 3,\cdot\cdot\cdot)$ as
follows: $\nu_{\alpha}
=~\sum_i
U_{\alpha i}\nu_i$. The $U_{\alpha i}$, the elements of the
Pontecorvo-Maki-Nakagawa-Sakata (PMNS) matrix represent the observable
mixing
angles in the basis where the charged lepton masses are diagonal.
In any other basis, one has $U =U^{\dagger}_{\ell}U_{\nu}$, where
the matrices $U_{\ell}$ and $U_{\nu}$ are the matrices that diagonalize
the charged lepton and neutrino mass matrices respectively. In the rest of
this talk, we will assume that neutrinos are Majorana particles; even
though this has not been experimentally established (and indeed, one of
the major goals of experimental efforts in neutrino physics is to confirm
or refute this), theoretical discussions are more convenient in this case
and also  most theoretical models predict neutrinos to be Majorana
particles.

For the case of three Majorana neutrinos, the PMNS matrix $U$ can be
written as: $VK$ where $V$ is the following matrix
\begin{equation}
\pmatrix{c_{12}c_{13} & s_{12}c_{13} & s_{13}e^{-i\delta} \cr
-s_{12}c_{23}-c_{12}s_{23}s_{13}e^{i\delta} &
c_{12}c_{23}-s_{12}s_{23}s_{13}e^{i\delta}& s_{23}c_{13} \cr
s_{12}s_{23}-c_{12}c_{23}s_{13}e^{i\delta} &
-c_{12}s_{23}-s_{12}c_{23}s_{13}e^{i\delta} & c_{23}c_{13} }
\end{equation}
and $K~=~diag(1, e^{i\phi_1},e^{i\phi_2})$. There are three angles and
three phases that characterize the mixings. In addition there the three
masses. It is the goal of the planned and current experiments to determine
 these 9 parameters as accurately as possible.  

In this review, I first summarize our present state of
understanding of these various parameters and then proceed to give an
overview of what kind of new physics is implied by these data.

\section{WHAT WE KNOW NOW:}

\subsection{Mass difference squares and Mixings}
Thanks to Super-Kamiokande results on the atmospheric neutrinos, the
results for solar neutrinos from Chlorine, Super-Kamiokande SAGE, GALLEX
 and most recently from the
SNO experiment on both charged and neutral currents,
as well as the KAMLAND, K2K and CHOOZ-PALO-Verde results, there now
appears to be a rough outline of the pattern of mixings among the various
neutrinos\cite{smirnov}. In terms of the parameters defined above, it
seems clear that both $\theta_{12}$ and
$\theta_{23}$ are large and the angle $\theta_{13}$ is small. This is in
sharp contrast with the corresponding mixings in the quark sector, which
are all small.

Oscillations also have given us the approximate values of the mass
difference squares for the neutrinos. The present
allowed values for the $\theta_{ij}$ as well as the $\Delta m^2$'s are (at
3$\sigma$
level):
$sin^22\theta_{23}\geq 0.92; 1.2\times 10^{-3} eV^2 \leq\Delta m^2_{23}
\leq 4.8\times 10^{-3} eV^2$; 
$0.70 \leq sin^22\theta_{12} \leq 0.95; 5.4\times 10^{-5} eV^2 \leq \Delta
m^2_{12} \leq 9.5\times 10^{-5} eV^2$;
$sin^2\theta_{13}\leq 0.23 $.
There is no information on any of the phases right now.

Since the oscillation data gives only the mass difference squares, it
allows for three possible arrangements of the different mass levels:
\begin{itemize}

\item (i) Normal hierarchy i.e. $m_1\ll m_2 \ll m_3$. In this case,
we can deduce the value of $m_3 \simeq \sqrt{\Delta m^2_{23}}\simeq
0.03-0.07$ eV. In this case $\Delta m^2_{23}\equiv m^2_3-m^2_2 > 0$.
 The solar neutrino oscillation involves the two lighter levels. The mass
of the lightest neutrino is unconstrained. If $m_1\ll m_2$, then we get
the value of $m_2 \simeq \simeq 0.008$ eV.

\item (ii) Inverted hierarchy i.e. $m_1 \simeq m_2 \gg m_3$ with
$m_{1,2} \simeq \sqrt{\Delta m^2_{23}}\simeq 0.03-0.07$ eV. In this case,
solar neutrino oscillation takes place between the heavier levels and we
have $\Delta m^2_{23}\equiv m^2_3-m^2_2 < 0$.

\item (iii) Degenerate neutrinos i.e. $m_1\simeq m_2 \simeq m_3$.

It is hoped that future long baseline experiments as well as the searches
for neutrinoless double beta decay can resolve between the different
possibilities. We will discuss this below. 

\end{itemize} 

\subsection{Overall scale for masses}
Oscillation experiments do not tell us about the overall scale of masses.
There are several ways to pin down this scale:

\subsubsection{Neutrino mass from beta decay}
 (i) One way is to directly search for the effect of nonzero
neutrino mass in the beta decay spectrum by looking for structure near the
end point of the electron energy spectrum. (A
commonly used nucleus is tritium.) It will measure a mass regardless of
whether neutrino is a Dirac or Majorana particle.
 In this case, one measures the
quantity $m_{\beta}\equiv \sqrt{\sum_i |U_{ei}|^2m^2_i}$. The Troitsk and
Mainz results put the present upper limit on $m_\beta \leq 2.2$
eV\cite{wil}. The proposed KATRIN experiment is projected to lower the
sensitivity down to 0.2 eV, which will have important implications for the
theory of neutrino masses. For instance, if the result is positive, it
will imply a degenerate spectrum; on the other hand a negative result will
be a very useful constraint.

\subsubsection{Neutrino mass from neutrinoless double beta decay}
 
 Second way is to search for neutrinoless double beta
decay, $\beta\beta_{0\nu}$, which can proceed if there is a sizeable
Majorana mass for the neutrino or if there are lepton number violating
interactions\cite{moh}. In the first case, one measures the quantity   
$m_{ee}~=~ \sum U^2_{ei} m_i$.  In the second case, the neutrino
necessarily has a
Majorana mass\cite{valle}; however, if the resulting induced mass is
small, then neutrinoless double beta decay will measure the strength of
the new interactions (such as doubly charged Higgs fields or
R-parity violating interactions etc ) rather than neutrino mass. There are
many
examples of models where new interactions can lead to  $\beta\beta_{0\nu}$
decay rate in the observable range without at the same time giving a
significant Majorana mass for the neutrinos. As a result, one must be
careful in interpreting any nonzero signal in a $\beta\beta_{0\nu}$
experiment and not jump to a conclusion that a direct measurement of
neutrino mass has been made. Way to tell whether such a nonzero
signal is neutrino mass or is a reflection of new interactions is to
supplement  $\beta\beta_{0\nu}$ decay results with collider searches for
the new interactions. Thus collider experiments such as those at LHC and
the double beta experiments play complementary role.

Present upper limits on  $\beta\beta_{0\nu}$ decay lifetimes from the
Heidelbergt-Moscow and can be translated to an upper limit on $m_{ee}\leq
0.3 $ eV or so. There is a claim of a discovery of   
neutrinoless   double beta decay in the enriched $^{76}Ge$ experiment by
the Heidelberg-Moscow collaboration\cite{klapdor}. Interpreted in terms of
a Majorana mass of the neutrino, this implies $m_{ee}$ between 0.11 eV to
0.56 eV. If confirmed, this result is of fundamental significance. There
have been much discussion of this in literature\cite{steve}.

\subsubsection{ Cosmology and neutrino mass}
 A very different way to get
information on the absolute scale of neutrino mass is from the study of
cosmic microwave radiation spectrum as well as the study of large scale
structure in the universe. A rough way this comes about is that if
neutrinos
are present in abundance in the universe at the epoch of structure
formation with a significant mass, it will affect structure formation. For
instance, for a given neutrino mass $m$, all structures on a scale smaller
than a value given by the inverse of neutrino mass will be washed away by
neutrino free streaming. This will reduce power on smaller scales. Thus
accurate information of the galaxy power spectrum for small scales can
help constrain neutrino mass. Recent results from the WMAP has put a limit
on the sum of neutrino masses $\sum m_i \leq 0.7-2$
eV\cite{wmap,hannestad}. More recdent results from SDSS sky survey has put
a limit of $\sum m_i \leq 1.6$ eV.  Hannestad\cite{hannestad} has
emphasized that these upper limits can change if there are more neutrino
species- e.g. for 5 neutrinos, $\sum m_i \leq 2.12$ eV if they are in
equlibrium at the epoch of BBN.

A point worth emphasizing is that the above result is valid for both a
Majorana and a Dirac neutrino.

These limits are already in the
interesting range: for instance if the limit $0.7$ eV is taken seriously,
it would imply that each individual neutrino must have an upper limit on
its mass of $0.23$ eV, which is same as the projected value from the
proposed KATRIN experiment. All these limits are going to be much smaller
once PLANCK satellite observations are carried out, thereby providing a
completely independent source of information on neutrino masses.
Furthermore, these results also have implications for models of sterile
neutrinos that attempt to explain the LSND results.

\subsection{Sterile neutrinos}
Another question of great importance in neutrino physics is the number of
neutrino species. Measurement of the invisible Z-width 
 in LEP-SLC experiments tell us that only three types of
 neutrinos couple to the W and Z boson. They correspond to the three known
neutrinos  $\nu_{e,\mu,\tau}$. This implies that if there are other
neutrino species, then they must
have little or no interaction with the W and Z. They are called sterile
neutrinos. So the question is: are there any sterile neutrinos and if
there are how many there are ?

\subsubsection{LSND and sterile neutrinos}
The first need for sterile neutrinos came from attempts to
explain\cite{caldwell}
Los Alamos Liquid Scintillation Detector (LSND) experiment\cite{lsnd} ,
where neutrino oscillations both from a stopped muon (DAR) as well as the
one accompanying the muon in pion decay
(known as the DIF) have apparently been observed. The evidence from the
DAR is statistically
more significant and is an oscillation from $\bar{\nu}_\mu$ to
$\bar{\nu}_e$. The mass and mixing parameter range that fits data is:
\begin{eqnarray}
 \Delta m^2 \simeq 0.2 - 2 eV^2; sin^22\theta \simeq 0.003-0.03
\end{eqnarray}
There are points at higher masses specifically at 6 eV$^2$ which are
also allowed by the present LSND data for small mixings.
KARMEN experiment at the Rutherford laboratory has very strongly
constrained the allowed parameter range of the LSND
data\cite{karmen}. Currently the
Miniboone experiment at Fermilab is under way to probe the LSND parameter
region\cite{louis}.

Since this  $\Delta m^2_{LSND}$ is so different from that  $\Delta
m^2_{\odot, A}$, the simplest way to explain these results is to
add one\cite{caldwell} or two\cite{sorel} sterile neutrinos. For the case
of one extra sterile neutrino, there are two scenarios: (i) 2+2 and
(ii) 3+1. In the first case, solar neutrino oscillation is supposed to be
from $\nu_e$ to $\nu_s$. This is ruked out by SNO neutral current data. In
the second case, one needs a two step process where $\nu_{\mu}$
undergoes indirect oscillation to $\nu_e$ due to a combined effect of 
$\nu_\mu-\nu_s$ and $\nu_e-\nu_s$ mixings (denoted by $U_{\mu,s}$ and
$U_{e s}$ respectively, rather than direct
$\nu_\mu-\nu_e$ mixing. As a result, the effective mixing angle in LSND is
given by $4U^2_{e s}U^2_{\mu s}$ and the measured mass difference is given
by that between 
 $\nu_{\mu,e}-\nu_s$ rather than  $\nu_\mu-\nu_e$. This scenario is
constrained by the fact that sterile neutrino
mixings are constrained by two sets of observations: one from the
accelerator searches for $nu_\mu$ and $\nu_e$ disappearance\cite{maltoni} 
and the second from big bang nucleosynthesis\cite{sarkar}.

The bounds on $U_{es}$ and $U_{\mu s}$ from accelerator experiments
such as Bugey, CCFR and CDHS are of course dependent on particular
value of $\Delta m^2_{\alpha s}$ but for a rough order of magnitude, we
have $U^2_{es}\leq 0.04$ for $\Delta m^2 \geq 0.1$ eV$^2$ and $U^2_{\mu s}
\leq 0.2$ for $\Delta m^2 \geq 0.4$ eV$^2$\cite{bilenky}.

It is worth pointing out that SNO neutral current data has ruled out pure
$\nu_e-\nu_s$ transition as an explanation of solar neutrino puzzle by
8$\sigma$'s; however, it still allows as much as 40\% admixture of sterile
neutrinos and as we will see below, the sterile neutrinos could very well
form a sub-dominant component in solar neutrino transitions.

 \subsubsection{BBN and sterile neutrinos}

Big bang nucleosynthesis (BBN) will put bounds on how many extra
neutrinos are allowed by the present observations of primordial
abundances of light elements. It is important to remember that
the mere
existence of a sterile neutrino does not conflict with BBN
results. It is effective only if its mass and mixings with active
neutrinos satisfy the constraint\cite{sarkar} 
\begin{equation}
\Delta m^2 sin^4 2\theta \geq \xi 10^{-5} eV^2.
\end{equation}
where $\xi$ is a number of order one and is flavor dependent. If for
example
one interprets the LSND result in terms of a sterile neutrino, one has
$sin^22\theta \simeq 10^{-3}$ and $\Delta m^2 \sim $ eV$^2$ which
satisfies the above condition and therefore such a sterile neutrino will
count as one extra species of neutrino. We therefore need to discuss
whether an extra neutrino species is allowed by light element production
at BBN epoch.

The light element abundances at BBN epoch depend on several factors such
as the baryon to photon ratio, chemical potential of the neutrinos which
measures the excess of neutrinos over anti-neutrinos (or lepton asymmetry
of the Universe) and the number of
neutrinos in equlibrium with radiation which determines the Hubble
expansion rate at that epoch. In generic models, one expects the lepton
asymmetry to be of order of the baryon asymmetry of the universe, in which
case it has no effect on BBN. Under this assumption,  one can
derive limits on $\Delta N_\nu$, the number of sterile neutrinos from
BBN, by using as input the primordial He$^4$ abundance (denoted as
$Y_p$) and the deuterium abundance $D/H$. The word ``primordial'' here is
crucial. Since the observed abundances may have undergone some
modification due to the age of the universe (e.g. stellar processing etc),
uncertainties can creep in when one derives the primordial abundances from
observed abundances. That this could be so for $Y_p$ has been noted by
many authors\cite{sarkar}. In particular, many people have noted that
$Y_p$ suffers from large systematic errors. The $D/H$ ratio on the 
other hand is believed to have less systematic uncertainties although it
is somewhat statistics limited. In any case, if both the presently
inferred values of 
$Y_p$ and $D/H$ are taken as inputs, the best fit point turns out to be
for $\Delta N_\nu\leq 0$\cite{lang} and the most likely value of
$\eta=5.7 \times 10^{-10}$. 

The WMAP experiment has now determined the
value of $\eta = (6.14\pm 0.25)\times 10^{-10}$. One may therefore take
this highly precise value of $\eta$ and try to combine it with the $D/H$
observations to constrain the number of neutrinos leaving aside the $Y_p$
value. This allows for as many as two extra neutrinos\cite{olive}. 

In any case, if the Helium data becomes more precise with less systematic
errors and there is independent evidence for sterile neutrinos, then this
would be a clear indication for different kind of new physics. One
possibility is there is a majoron\cite{cmp} coupled to neutrinos that
contributes significantly to matter effecty at the era of BBN to suppress
the sterile to active neutrino mixing\cite{roth}. This can give rise to a
plethora of new phenomena of both laboratory and astrophysical
interest\cite{pak}.

Let us briefly discuss the implications of this discussion for
interpretation of the LSND results in terms of sterile neutrinos.
We remind the reader that there are three possible sterile neutrino
scenarios for LSND: (i) 2+2\cite{caldwell}; (ii) 3+1\cite{barger} and
(iii) 3+2\cite{sorel}.
It appears that 2+2 models are disfavored by a combination of accelerator
as well as solar and atmospheric neutrino data. The 3+1 scenario is
however marginally allowed for only specific mass and mixing values. The
only one that is consistent with the WMAP data is the one with $\Delta
m^2=0.8$ eV$^2$ and $sin^22\theta = 2\times 10^{-3}$. On the other hand
the 3+2 scenario requires two sterile neutrinos one of which has a mass
around 1 eV and a second one around $4.5$ eV. Both have mixings with
$\nu_{e,\mu}$ which bring them to equilibrium at the epoch of
nucleosynthesis. This scenario would then appear to be in conflict with
mass bounds on $\sum m_i $ from WMAP\cite{hannestad}.

Recently, another possibility for the existence a sterile neutrino has
been suggested in ref.\cite{deho}. It was noted in this paper that the 
now favored LMA solution to the solar neutrino puzzle runs into two
possible difficulties: (i) it predicts the Argon production rate which
higher than observations at 2$\sigma$ level (LMA prediction is 3 SNU's
as against the observed value of $2.56\pm 0.23$) and (ii) SNO data does
not show a rise in the low energy region that is predicted by the LMA
solution. Both these difficulties can be resolved if there is a
sterile neutrino with $\nu_e-\nu_s$ $\Delta m^2\simeq (0.2-2)\times
10^{-5}$ eV$^2$ and mixing angle $sin^22\alpha \simeq 10^{-3}-10^{-5}$.
Such a sterile neutrino escapes all the above cosmological bounds and is
therefore quite acceptable.

\subsection{CP violation}
It is clear from Eq. (1) that for Majorana neutrinos, there are three CP
phases that characterize neutrino mixings and a complete understanding of
leptonic mixing will be incomplete without a knowledge of these phases.
There are two possible ways to explore CP phases: (i) one way is to via
the long baseline experiments and looking for differences between neutrino
and anti-neutrino survival probabilities\cite{marciano}; (ii) another way
is to use possible connection with cosmology. It has often been argued
that neutrinoless double beta decay may also provide a alternative way to
explore CP violation. A detailed discussion of these issues is beyond the
scope of this article. For some discussions, see \cite{andre}.

\section{IMPLICATIONS FOR PHYSICS BEYOND THE STANDARD MODEL:}
These discoveries involving neutrinos, which have provided the
first evidence for physics beyond the standard model, have raised a
number of challenges for theoretical physics.
Foremost among them are, (i) an  understanding of the smallness of
neutrino masses and (ii) understanding the vastly different pattern of
mixings among neutrinos from the quarks. Specifically, a key question
is whether it is possible
to reconcile the large neutrino mixings with small quark mixings in grand
unified frameworks suggested by supersymmetric gauge coupling
unifications that unify quarks and leptons.

\subsection{Seesaw mechanism for small neutrino masses: type I and type
II seesaw}
The first challenge posed by neutrinos, i.e. the extreme smallness of
neutrino masses is elegantly answered by the seesaw mechanism
\cite{seesaw1}
which requires an extension of the standard model that includes heavy
right handed neutrinos. The light neutrino mass matrix obtained by
integrating out heavy right-handed neutrinos is given by
\begin{equation}
{M}_\nu = - M_{\nu}^D M_R^{-1} (M_\nu^D)^T,
\end{equation}
where $M_\nu^D$ is the Dirac neutrino mass matrix and $M_R$ is the
right-handed Majorana mass matrix.

There are several reasons why the seesaw mechanism is
appealing: (i) first, it
restores quark-lepton symmetry to the standard model; (ii) secondly, it
allows B-L to be an anomaly free gaugeable symmetry, thereby expanding
the minimal electroweak gauge group to the the left-right symmetric group
$SU(2)_L\times SU(2)_R\times U(1)_{B-L}\times SU(3)_c$, which is known to
provide a new way to understand the observed parity violation in weak
interactions. In the left-right symmetric model, the electric charge
formula is given by\cite{marshak}:
\begin{equation}
Q~=~I_{3L}+I_{3R}+\frac{B-L}{2}
\end{equation}
First of all unlike the corresponding formula in the standard model,, this
formula involves only physical quantities like weak isospin and B-L
quantum numbers. Secondly,
taking variation of both sides of this charge equation above the weak
scale, we
get $\Delta I_{3R}\simeq- \frac{\Delta(B-L)}{2}$. For purely leptonic
processes since $\Delta B=0$ and weak interactions have parity
violation, one must have lepton number violation. In particular, it
implies that neutrino in this theory is
naturally a Majorana particle. The presence of the heavy right handed
neutrino  also opens up a new way to understand the origin of matter in
the universe from baryogenesis via leptogenesis arising from the decay of
the right handed neutrinos in combination with CP violation.

 The above
formula for the neutrino mass matrix is called type I seesaw formula.
The right-handed Majorana mass scale, $M_R$, is more or less determined
by the mass squared difference needed to understand the atmospheric
neutrino data to be around $10^{14}$ GeV, (if we assume that the Dirac
neutrino mass is same as up-type quark mass). Note that this scale is
tantalizingly close to the GUT scale suggesting that
grand unifed theories may provide a natural framework to unravel the
mysteries of neutrino physics. In this talk I will strengthen this
argument by providing a simple model where this happens.
The seesaw formula in Eq. (1)  also implies that the scale at which the
left-right symmetric gauge group manifests itself is near the GUT scale.
One must then look for a GUT group that contains $SU(3)_c\times
SU(2)_L\times SU(2)_R\times U(1)_{B-L}$. SO(10) happens to be the minimal
such group. We would therefore explore to what extent we can understand
the properties of the neutrinos within the SO(10) group.

Before proceeding further, let us note a general phenomenon that when
the theory containing the $N_R$ becomes parity symmetric, the seesaw
formula changes and becomes:
\begin{equation}
{ M}_\nu^{\rm II} = M_L - M_\nu^D M_R^{-1} (M_\nu^D)^T,
\end{equation}
where $M_L = f v_L$ and $M_R=f v_R$, where $v_{L,R}$ are the vacuum
expectation values of Higgs fields that couple to the right and lefthanded
neutrinos.
This formula for the neutrino mass matrix is called type II seesaw formula
\cite{seesaw2}.

\subsection{Understanding large mixings for degenerate neutrinos}

A major puzzle of quark lepton physics is the diverse nature of the
mixing angles between quarks and leptons. Whereas in the quark sector the
mixing angles are small, for the leptons they are large. 

In order to understand the mixing angles, we have to study the mass
matrices for the charged leptons and neutrinos. Since we can choose an
arbitrary basis for either the charged leptons or the neutrinos without
effecting weak interactions, we will work in a basis where charged lepton
mass matrix is diagonal. One can then look for the types of mass matrices
for neutrinos that can lead to bi-large mixings and try to understand
them in terms of new physics.

\noindent (i) {\it The case of normal hierarchy:}
A neutrino mass matrix that leads to bi-large mixing in this case has the
form:
\begin{equation}
M_\nu~=~m_0\pmatrix{\epsilon & \epsilon & \epsilon\cr \epsilon &
1+\epsilon &
1\cr \epsilon & 1 & 1}
\end{equation}
where $m_0$ is $\sqrt{\Delta m^2_{ATM}}$. We have omitted order one
coefficients in front of the $\epsilon$'s. This leads to $tan
\theta_A\simeq 1$, $\Delta m^2_{\odot}/\Delta m^2_A \simeq \epsilon^2$ and 
also large solar angle. For the LMA I solution, we find the interesting
result that $\epsilon \sim \lambda$ where $\lambda$ is the Cabibbo angle
($\simeq 0.22$). This  could be a signal of hidden quark lepton
connection.
In fact we will see below that in the context of a minimal SO(10) model,
this connection is realized in a natural manner.

\noindent(ii) {\it The case of inverted hierarchy:}
The elements of the neutrino mass matrix in this case have a slightly
different pattern.
\begin{eqnarray}
{ M}_\nu=m_0~\left(\begin{array}{ccc} \epsilon &
c & s\\ c & \epsilon & \epsilon\\ s & \epsilon &
\epsilon\end{array}\right).
\end{eqnarray}
where $c=cos\theta$ and $s=sin\theta$ and it denotes the atmospheric
neutrino mixing angle. An interesting point about this mass matrix is that
in the limit of $\epsilon\rightarrow 0$, it has $L_e-L_\mu-L_\tau$
symmetry. One therefore might hope that if inverted hierarchy structure is
confirmed, it may provide evidence for this leptonic symmetry and which
can be an important clue to new physics beyond the standard model.
However the fact that the solar mixing angle appears to be far from
being maximal means that  $L_e-L_\mu-L_\tau$ symmetry must be badly
broken.

\noindent (iii) {\it Degenerate neutrinos:}
In this case, there are two ways to proceed: one may add the unit matrix
to either of the above mass matrices to understand large mixings or look
for some dynamical ways by which large mixings can arise. It
turns that if neutrinos are mass degenerate, one can generate large
mixings out of small mixings\cite{babu1,balaji} purely as a consequence
of radiative corrections. We will call this possibility radiative
magnification.

Let us illustrate the basic mechanism for the case of two generations.
The mass matrix in the $\nu_\mu-\nu_\tau$
sector\cite{balaji} cab written in the flavor basis as:
\begin{eqnarray}
{M_F(M_R)} =  U(\theta)
       \left(\begin{array}{cc} m_1 & 0 \\ 0 & m_2 \end{array}
\right) U(\theta)^{\dagger}
\label{uudag}
\end{eqnarray}
where $U(\theta)~=  \left(\begin{array}{cc} C_\theta & S_\theta \\
-S_\theta &
C_\theta \end{array} \right)$.
 This mass matrix is defined at the
seesaw (GUT) scale, where we assume the mixing angles to be small. As we
extrapolate this mass matrix down to
the weak scale, radiative corrections modify it to the form\cite{many}
\begin{eqnarray}
\cal{M_F (M_Z)} ~=~ \cal{R}\cal{M_F (M_R)}\cal{R}
\label{mf-fin}
\end{eqnarray}
where $\cal{R}~=~ \left(\begin{array}{cc} 1+\delta_\mu & 0 \\ 0 &
1+\delta_\tau \end{array} \right)$. Note that $\delta_{\mu} \ll
\delta_\tau$. So if we ignore $\delta_\mu$, we find that the $\tau\tau$
entry of the $\cal{M_F(M_Z)}$ is changed compared to its value at the
seesaw scale. If the seesaw scale mass eigenvalues are sufficiently close
to each other, then the two eigenvalues of the neutrino mass matrix at the
$M_Z$ scale can be same leading to maximal mixing (much like MSW matter
resonance effect) regardless what the values of the mixing angles at the
seesaw scale are. Thus at the seesaw scale can even be same as the quark
mixing angles as a quark-lepton symmetric theory would require. We call
this phenomenon radiative magnification of mixing angles. It requires no
assumption other than the near degeneracy of neutrino mass eigenvalues and
is a new way to understand large mixings.

This has recently been generalized to the case of three
neutrinos\cite{parida}, where assuming the neutrino mixing angles at the
seesaw scale to be same as the quark mixing angles renormalization group
extrapolation alone leads to large solar and atmospheric as well as small
$\theta_{13}$ at the weak scale provided the common mass of the neutrinos 
$m_0\geq 0.1$ eV.
We find that while both the solar and atmospheric mixing angles become
large, the $\theta_{13}$ parameter remains small ($0.08$). 

An important
prediction of this model is that the common mass of the neutrinos must be
bigger than 0.1 eV, as already noted for the radiative magnification
mechanism to work. This
result can be tested in the proposed neutrinoless double beta decay
experiments. It is within the range of values reported in
ref.\cite{klapdor}.

\section{MINIMAL SO(10)GRAND UNIFICATION AND NEUTRINO MIXINGS:}

The minimal grand unification model for neutrinos is the one
based on the SO(10) group since all standard model fermions and
the right-handed neutrino fit into
the $\mathbf{16}$-dimensional representaion of SO(10),
resulting not only in a complete unification of the quarks and leptons but
also yielding possible relations between the quark and lepton mass
matrices.
One may therefore hope that the neutrino oscillation parameters might be
predictable in an SO(10) theory.

There are two simple routes to realistic SO(10) model building. In the
first class, one may have smaller representations for the Higgs fields
like {\bf 10} and {\bf 16} multiplets\cite{other}. In this case, one must
necessarily
introduce
nonrenormalizable terms to the superpotential to implement the seesaw
they break R-parity which then induces rapid proton decay at an
unacceptable level.

An alternative is to introduce both {\bf 10} and {\bf 126} Higgs
multiplets to give fermion masses. In this class of models, there is no
need to invoke nonrenormalizable terms and also R-parity is an automatic
symmetry of the model. This naturally prevents the baryon and lepton
number
violating terms that give rise to rapid proton decay and also
guarantees a naturally stable supersymmetric dark matter. It is these
class of minimal models\cite{babu,goran,goh} that we discuss here.

In SO(10) models of this type,
the {\bf 126} multiplet contains two parity partner Higgs submultiplets
(called $\Delta_{L,R}$) which couple to $\nu_L\nu_L$ and $N_RN_R$
respectively and after spontaneous symmetry breaking lead to the type II
seesaw formula for neutrinos, which plays an important role in magnifying
the neutrino mixings despite quark-lepton unification\cite{goran,goh}.

As we will see a further advantage of using {\bf 126} multiplet is that it
unifies the charged fermion Yukawa couplings with
the couplings that contribute to righthanded as well as lefthanded
neutrino masses, as long as we do not include nonrenormalizable
couplings in the superpotential. This can be seen as
follows\cite{babu}: it is the set {\bf 10}+${\bf
\overline{126}}$ out of which the MSSM Higgs doublets emerge; the
later also contains the multiplets $(3,1,10)+(1,3,\overline{10})$
which are responsible for not only lefthanded but also the right
handed neutrino masses in the type II seesaw formula.
Therefore all fermion masses in the model are arising
from only two sets of $3\times 3$ Yukawa matrices one denoting the
{\bf 10} coupling and the other denoting ${\bf \overline{126}}$
couplings.

The SO(10) invariant superpotential giving the Yukawa couplings of the
{\bf
16} dimensional matter spinor $\psi_i$ (where $i,j$ denote generations)
with the Higgs fields $H_{10}\equiv
{\bf 10}$ and $\Delta\equiv {\bf \overline{126}}$.
\begin{eqnarray}
{W}_Y &=&  h_{ij}\psi_i\psi_j H_{10} + f_{ij} \psi_i\psi_j\Delta
\end{eqnarray}
 In terms of the GUT scale Yukawa couplings, one can write the
fermion mass matrices (defined as ${ L}_m~=~\bar{\psi}_LM\psi_R$) at
the seesaw scale as:
\begin{equation}
M_u = \bar{h} + \bar{f};
\end{equation}
\begin{equation}
M_d = \bar{h}r_1 + \bar{f}r_2;
\end{equation}
\begin{equation}
M_e = \bar{h}r_1 -3r_2 \bar{f};
\end{equation}
\begin{equation}
M_{\nu^D} = \bar{h} -3 \bar{f}.
\end{equation}
where $\bar{h}$, $ \bar{f}$, $r_{1,2}$ are functions of GUT scale
Yukawa couplings and mixing parameters discussed.
These mass sumrules provide the first
important ingredient in discussing the neutrino sector.
 To see this let us note that they lead to the following sumrule
involving the
charged lepton, up and down quark masses:
\begin{equation}\label{main}
    k \tilde{M}_l=r \tilde{M}_d+\tilde{M}_u
\end{equation} where $k$ and $r$ are
 functions of the symmetry breaking parameters of the model. It is clear
from the above equation that smallquark mixings imply that the
contribution the charged leptons to the neutrino mixing matrix
i.e. $U_{\ell}$ in the formula $U_{PMNS}~=~U^{\dagger}_{\ell}U_{\nu}$
is close to identity and the entire contribution therefore comes from
$U_\nu$. Below we show that $U_nu$ has the desired form with
$\theta_{12}$ and $\theta_{23}$ large and $\theta_{13}$ small.

\subsection{Maximal neutrino mixings from type II seesaw}
In order to see how the type II seesaw formula provides a simple way to
understand large neutrino mixings in this model, note that in certain
domains of the parameter space of the model, the second matrix in the type
II seesaw formula can much smaller than the first term. This can happen
for instance when $V_{B-L}$ scale is much higher than $10^{16}$ GeV. When
this happens, one can derive the sumrule
\begin{eqnarray}
{ M}_{\nu} &=& a(M_{\ell}-M_d)
\label{key}\end{eqnarray}
This equation is key to our discussion of the neutrino masses and mixings.

Using Eq. (\ref{key}) in second and third generation sector, one
can understand how large mixing angle emerges.

Let us first consider the two generation case \cite{goran}. The known
hierarchical
structure of quark and lepton
masses as well as the known small mixings for quarks suggest that
the matrices $M_{\ell,d}$ for the second and third generation have

\begin{eqnarray}
M_{\ell}~\approx~m_\tau\left(\begin{array}{cc}\lambda^2
&\lambda^2\\
\lambda^2 & 1\end{array}\right)\\ \nonumber
M_q ~\approx~m_b \left(\begin{array}{cc}\lambda^2 & \lambda^2\cr \lambda^2
& 1\end{array}\right)
\end{eqnarray}
where $\lambda \sim 0.22$ (the Cabibbo angle).
 It is well known that in supersymmetric
theories, when low energy quark and lepton masses are extrapolated
to the GUT scale, one gets approximately that $m_b\simeq m_\tau$.
One then sees from the above sumrule for neutrino masses Eq.
(\ref{key}) that there is a cancellation in the $(33)$ entry of the
neutrinomass matrix and all entries are of
same order $\lambda^2$ leading very naturally to the atmospheric
mixing angle to be large. Thus one has a natural understanding of
the large atmospheric neutrino mixing angle. No extra symmetries
are assumed for this purpose.

For this model to be a viable one for three generations, one must
show that the same $b-\tau$ mass convergence at GUT scale also
explains the large solar angle $\theta_{12}$ and a small
$\theta_{13}$. This has been demonstrated in a recent
paper\cite{goh}.

To see how this comes about, note
that in the basis where the down
quark mass matrix is diagonal, all the quark mixing effects are
then in the up quark mass matrix i.e. $M_u ~=~ U^T_{CKM}M^d_u
U_{CKM}$. Using the Wolfenstein parametrization for quark mixings,
we can conclude that that we have
\begin{eqnarray}
M_{d}~\approx ~m_{b}\left(\begin{array}{ccc}\lambda^4 & \lambda^5
&\lambda^3\\ \lambda^5 & \lambda^2& \lambda^2 \\ \lambda^3 & \lambda^2 &
1\end{array}\right)
\end{eqnarray}
and $M_{\ell}$ and $M_d$ have roughly similar pattern due to the
sum rule . In the above equation, the matrix elements are
supposed to give only the approximate order of magnitude. As we
extrapolate the quark masses to the GUT scale, due to the fact
just noted i.e. $m_b-m_\tau \approx m_{\tau}\lambda^2$,
 the neutrino mass matrix $M_\nu~=~c(M_d-M_\ell)$
takes roughly the form:
 \begin{eqnarray}
M_{\nu}~=~c(M_d-M_\ell)\approx ~m_0\left(\begin{array}{ccc}\lambda^4 &
\lambda^5
&\lambda^3\\ \lambda^5 & \lambda^2 & \lambda^2 \\ \lambda^3 & \lambda^2
& \lambda^2\end{array}\right)
\end{eqnarray}
 It is then easy to see from this mass matrix that both the $\theta_{12}$
(solar angle) and $\theta_{23}$ (the atmospheric angle) are
large. It also turns out that the ratio of masses $m_2/m_3\approx \lambda$
which explains the milder hierarchy among neutrinos compared to
that among quarks. Furthermore, $\theta_{13}\sim \lambda$.
A detailed numerical analysis for this modelhas been carried out in
\cite{goh} and it substantiates the above analytical reasoning and
makes detailed predictions for the mixing angles\cite{goh}. We find that
the predictions for $sin^22\theta_{\odot}\simeq 0.9-0.94$,
$sin^22\theta_A \leq 0.92$, $\theta_{13}\sim 0.16$ and $\Delta
m^2_{\odot}/\Delta m^2_A\simeq 0.025-0.05$ are all in agreement with
data. Furthermore the prediction for $\theta_{13}$ is in a range that can
be tested partly in the MINOS experiment but more completely in the 
proposed long baseline experiments.

In conclusion, the progress in the field of neutrino mass has been
phenomenal. A lot is now known about the masses and mixings but a lot of
crucial informations are still missing (e.g. whether the neutrino is a
Dirac or Majorana particle; sign of $\Delta m^2_{23}$ to name a
few). These need to be probed so that finally one can say that we know as
much about the leptons as we do about quarks. As far as their implications
for new physics beyond the standard model are concerned, seesaw mechanism
and the existence of the right handed neutrino are emerging as two 
dominant ideas but asto the details of mixings, no clear solution has
emerged yet. The introduction of the right handed neutrino not only makes
the particle physics quark-lepton symmetric but it also makes the weak
interactions asymptotically parity conserving. The possibility that
neutrino mass owes its origin to grand unification is a tantalizing
possibility. The next decade promises to be as exciting in neutrino
physics as the past one.

\end{document}